\documentclass[twocolumn,aps,prc,superscriptaddress]{revtex4}

\usepackage{sidecap}
\usepackage{ulem}
\usepackage{epsfig}
\usepackage{amsmath,amssymb,amsthm}
\usepackage{graphicx}
\usepackage{bm}
\usepackage{color,soul}
\usepackage{url}

\DeclareMathAlphabet{\mathpzc}{OT1}{pzc}{m}{it}

\begin{document}

\renewcommand{\textfraction}{0.00}


\newcommand{\vAi}{{\cal A}_{i_1\cdots i_n}}
\newcommand{\vAim}{{\cal A}_{i_1\cdots i_{n-1}}}
\newcommand{\vAbi}{\bar{\cal A}^{i_1\cdots i_n}}
\newcommand{\vAbim}{\bar{\cal A}^{i_1\cdots i_{n-1}}}
\newcommand{\htS}{\hat{S}}
\newcommand{\htR}{\hat{R}}
\newcommand{\htB}{\hat{B}}
\newcommand{\htD}{\hat{D}}
\newcommand{\htV}{\hat{V}}
\newcommand{\cT}{{\cal T}}
\newcommand{\cM}{{\cal M}}
\newcommand{\cMs}{{\cal M}^*}
\newcommand{\vk}{\vec{\mathbf{k}}}
\newcommand{\bk}{\bm{k}}
\newcommand{\kt}{\bm{k}_\perp}
\newcommand{\kp}{k_\perp}
\newcommand{\km}{k_\mathrm{max}}
\newcommand{\vl}{\vec{\mathbf{l}}}
\newcommand{\bl}{\bm{l}}
\newcommand{\bK}{\bm{K}}
\newcommand{\bb}{\bm{b}}
\newcommand{\qm}{q_\mathrm{max}}
\newcommand{\vp}{\vec{\mathbf{p}}}
\newcommand{\bp}{\bm{p}}
\newcommand{\vq}{\vec{\mathbf{q}}}
\newcommand{\bq}{\bm{q}}
\newcommand{\qt}{\bm{q}_\perp}
\newcommand{\qp}{q_\perp}
\newcommand{\bQ}{\bm{Q}}
\newcommand{\vx}{\vec{\mathbf{x}}}
\newcommand{\bx}{\bm{x}}
\newcommand{\tr}{{{\rm Tr\,}}}
\newcommand{\bc}{\textcolor{blue}}

\newcommand{\beq}{\begin{equation}}
\newcommand{\eeq}[1]{\label{#1} \end{equation}}
\newcommand{\ee}{\end{equation}}
\newcommand{\bea}{\begin{eqnarray}}
\newcommand{\eea}{\end{eqnarray}}
\newcommand{\beqar}{\begin{eqnarray}}
\newcommand{\eeqar}[1]{\label{#1}\end{eqnarray}}

\newcommand{\half}{{\textstyle\frac{1}{2}}}
\newcommand{\ben}{\begin{enumerate}}
\newcommand{\een}{\end{enumerate}}
\newcommand{\bit}{\begin{itemize}}
\newcommand{\eit}{\end{itemize}}
\newcommand{\ec}{\end{center}}
\newcommand{\bra}[1]{\langle {#1}|}
\newcommand{\ket}[1]{|{#1}\rangle}
\newcommand{\norm}[2]{\langle{#1}|{#2}\rangle}
\newcommand{\brac}[3]{\langle{#1}|{#2}|{#3}\rangle}
\newcommand{\hilb}{{\cal H}}
\newcommand{\pleft}{\stackrel{\leftarrow}{\partial}}
\newcommand{\pright}{\stackrel{\rightarrow}{\partial}}
\newcommand{\dif}{\mathrm{d}}
\newcommand{\pT}{p_\perp}
\newcommand{\sNN}{\sqrt{s_{\mathrm{NN}}}}
\newcommand{\RAA}{R_{\mathrm{AA}}}

\title{Shape of the quark gluon plasma droplet reflected in the high-$\pT$ data}

\author{Magdalena Djordjevic\footnote{E-mail: magda@ipb.ac.rs}}
\affiliation{Institute of Physics Belgrade, University of Belgrade, Serbia}

\author{Stefan Stojku}
\affiliation{Institute of Physics Belgrade, University of Belgrade, Serbia}

\author{Marko Djordjevic}
\affiliation{Faculty of Biology, University of Belgrade, Serbia}

\author{Pasi Huovinen}
\affiliation{Institute of Physics Belgrade, University of Belgrade, Serbia}

\begin{abstract}
  We show, through analytic arguments, numerical calculations, and
  comparison with experimental data, that the ratio of the high-$\pT$
  observables $v_2/(1-\RAA)$ reaches a well-defined saturation value
  at high $p_\perp$, and that this ratio depends only on the spatial
  anisotropy of the quark gluon plasma (QGP) formed in
  ultrarelativistic heavy-ion collisions. With expected future
  reduction of experimental errors, the anisotropy extracted from
  experimental data will further constrain the calculations of initial
  particle production in heavy-ion collisions and thus test our
  understanding of QGP physics.
\end{abstract}


\maketitle

{\it Introduction:} The major goal of relativistic heavy-ion
physics~\cite{QGP1,QGP2,QGP3,QGP4} is understanding the properties of
the new form of matter called quark-gluon plasma
(QGP)~\cite{Collins,Baym}, which, in turn, allows understanding
properties of QCD matter at its most basic level. Energy loss of rare
high-momentum partons traversing this matter is known to be an
excellent probe of its properties. Different observables such as the
nuclear modification factor $\RAA$ and the elliptic flow parameter
$v_2$ of high-$\pT$ particles, probe the medium in different manners,
but they all depend not only on the properties of the medium, but also
on the density, size, and shape of the QGP droplet created in a
heavy-ion collision. Thus drawing firm conclusions of the material
properties of QGP is very time consuming and requires simultaneous
description of several observables. It would therefore be very useful
if there were an observable, or combination of observables, which
would be sensitive to only one or just a few of all the parameters
describing the system.

For high-$\pT$ particles, spatial asymmetry leads to different paths,
and consequently to different energy losses. Consequently, $v_2$
(angular differential suppression) carries information on both the
spatial anisotropy and material properties that affect energy loss
along a given path. On the other hand, $\RAA$ (angular average
suppression) carries information only on material properties affecting
the energy loss~\cite{DREENAc,DREENAb,Thorsten,Molnar}, so one might
expect to extract information on the system anisotropy by taking a
ratio of expressions which depend on $v_2$ and $\RAA$. Of course, it
is far from trivial whether such intuitive expectations hold, and what
combination of $v_2$ and $\RAA$ one should take to extract the spatial
anisotropy. To address this, we here use both analytical and numerical
analysis to show that the ratio of $v_2$ and $1-\RAA$ at high $\pT$
depends only on the spatial anisotropy of the system. This approach
provides a complementary method for evaluating the anisotropy of the
QGP fireball, and advances the applicability of high-$p_\perp$ data to
a new level as, up to now, these data were mainly used to study the
jet-medium interactions, rather than inferring bulk QGP parameters.

{\it Anisotropy and high-$\pT$ observables:} In~\cite{DREENAb,NewObservable},
we showed that at very large values
of transverse momentum $\pT$, the fractional energy loss $\Delta E/E$
(which is very complex, both analytically and numerically, due to
inclusion of multiple effects, see {\it Numerical results} for more
details) shows asymptotic scaling behavior
\begin{eqnarray}
 \Delta E/E \approx \chi (\pT) \langle T \rangle^a \langle L \rangle^b,
 \label{ElossEstimate}
\end{eqnarray}
where $\langle L \rangle$ is the average path length traversed by the
jet, $\langle T \rangle$ is the average temperature along the path of
the jet, $\chi$ is a proportionality factor (which depends on initial
jet $p_\perp$), and $a$ and $b$ are proportionality factors which
determine the temperature and path-length dependence of the energy
loss. Based on Refs.~\cite{BDMPS,ASW,GLV,HT}, we might expect values like
$a=3$ and $b=1$ or $2$, but a fit to a full-fledged calculation yields
values $a \approx 1.2$ and $b \approx 1.4$~\cite{NewObservable,MD_5TeV}.
Thus the temperature dependence of the energy loss is close to
linear, while the length dependence is between linear and
quadratic. To evaluate the path length we follow
Ref.~\cite{ALICE_charm}:
\begin{equation}
  L(x,y,\phi) = \frac{\int_0^\infty\dif\lambda\,\lambda\,
                                  \rho(x+\lambda\cos(\phi),y+\lambda\sin(\phi))}
                     {\int_0^\infty\dif\lambda\,
                                  \rho(x+\lambda\cos(\phi),y+\lambda\sin(\phi))},
\end{equation}
which gives the path length of a jet produced at point $(x,y)$ heading
to direction $\phi$, and where $\rho(x,y)$ is the initial density
distribution of the QGP droplet. To evaluate the average path length
we take average over all directions and production points.

If $\Delta E/E$ is small (i.e., for high $p_\perp$ and in peripheral
collisions), we obtain~\cite{DREENAc,DREENAb,NewObservable}
\begin{equation}
 \RAA \approx 1-\xi \langle T \rangle^a \langle L \rangle^b,
 \label{RaaEstimate}
\end{equation}
where $\xi = (n-2)\chi/2$, and $n$ is the steepness of a power law
fit to the transverse momentum distribution, $\dif N/\dif\pT \propto 1/\pT^n$.
Thus $1-\RAA$ is proportional to the average size and temperature of
the medium. To evaluate the anisotropy we define the average path lengths in
the in-plane and out-of-plane directions,
\begin{eqnarray}  \label{LinLout}
  \langle L_{in}\rangle
  & = & \frac{1}{\Delta\phi}\int_{-\Delta\phi/2}^{\Delta\phi/2}\!\!\!\!\dif\phi\,\langle L(\phi)\rangle \\
  \langle L_{out}\rangle                                                \nonumber
  & = & \frac{1}{\Delta\phi}\int_{\pi/2-\Delta\phi/2}^{\pi/2+\Delta\phi/2}\!\!\!\!\dif\phi\,\langle L(\phi)\rangle,
\end{eqnarray}
where $\Delta\phi = \pi/6$~\cite{PHENIX} is the acceptance angle with
respect to the event plane (in-plane) or orthogonal to it
(out-of-plane), and $\langle L(\phi)\rangle$ the average path length
in $\phi$ direction.  Note that the obtained calculations are robust
with respect to the precise value of the small angle $\pm \Delta\phi/2$,
but we still keep a small cone $(\pm \pi/12)$ for $\RAA^{in}$ and
$\RAA^{out}$ calculations, to have the same numerical setup as in our
Ref.~\cite{DREENAb}. Now we can write
$\langle L\rangle = (\langle L_{out}\rangle + \langle L_{in}\rangle)/2$
and $\Delta L = (\langle L_{out} \rangle - \langle L_{in}\rangle)/2 $.
Similarly, the average temperature along the path length can be split
to average temperatures along paths in in- and out-of-plane
directions, $\langle T_{in}\rangle = \langle T\rangle + \Delta T$ and
$\langle T_{out}\rangle = \langle T\rangle - \Delta T$. When applied
to an approximate way to calculate $v_2$ of high-$\pT$
particles~\cite{Christiansen:2013hya}, we obtain~\footnote{Note that
  the first approximate equality in Eq.~(\ref{v2approx}) can be shown
  to be exact if the higher harmonics $v_4$, $v_6$, etc., are zero,
  and the opening angle where $\RAA^{in}$ and $\RAA^{out}$ are
  evaluated is zero (cf.~definitions of $\langle L_{out}\rangle$ and
  $\langle L_{in}\rangle$, Eq.~(\ref{LinLout})).}
\begin{eqnarray}
  v_{2} & \approx & \frac{1}{2} \frac{\RAA^{in} - \RAA^{out}}{\RAA^{in} + \RAA^{out}}
          \approx \frac{\xi \langle T_{out}\rangle^a \langle L_{out}\rangle^b
                      - \xi \langle T_{in} \rangle^a \langle L_{in}\rangle^b}{4} \nonumber \\
        & \approx & \xi \langle T\rangle^a \langle L\rangle^b
                     \left( \frac{b}{2} \frac{\Delta L}{\langle L \rangle}
                           - \frac{a}{2}\frac{\Delta T}{\langle T \rangle} \right),
    \label{v2approx}
\end{eqnarray}
where we have assumed that $\xi \langle T \rangle^a \langle L \rangle^b \ll 1$, and that
$\Delta L/\langle L\rangle $ and $\Delta T/\langle T\rangle $ are small as well.

\begin{figure}
\epsfig{file=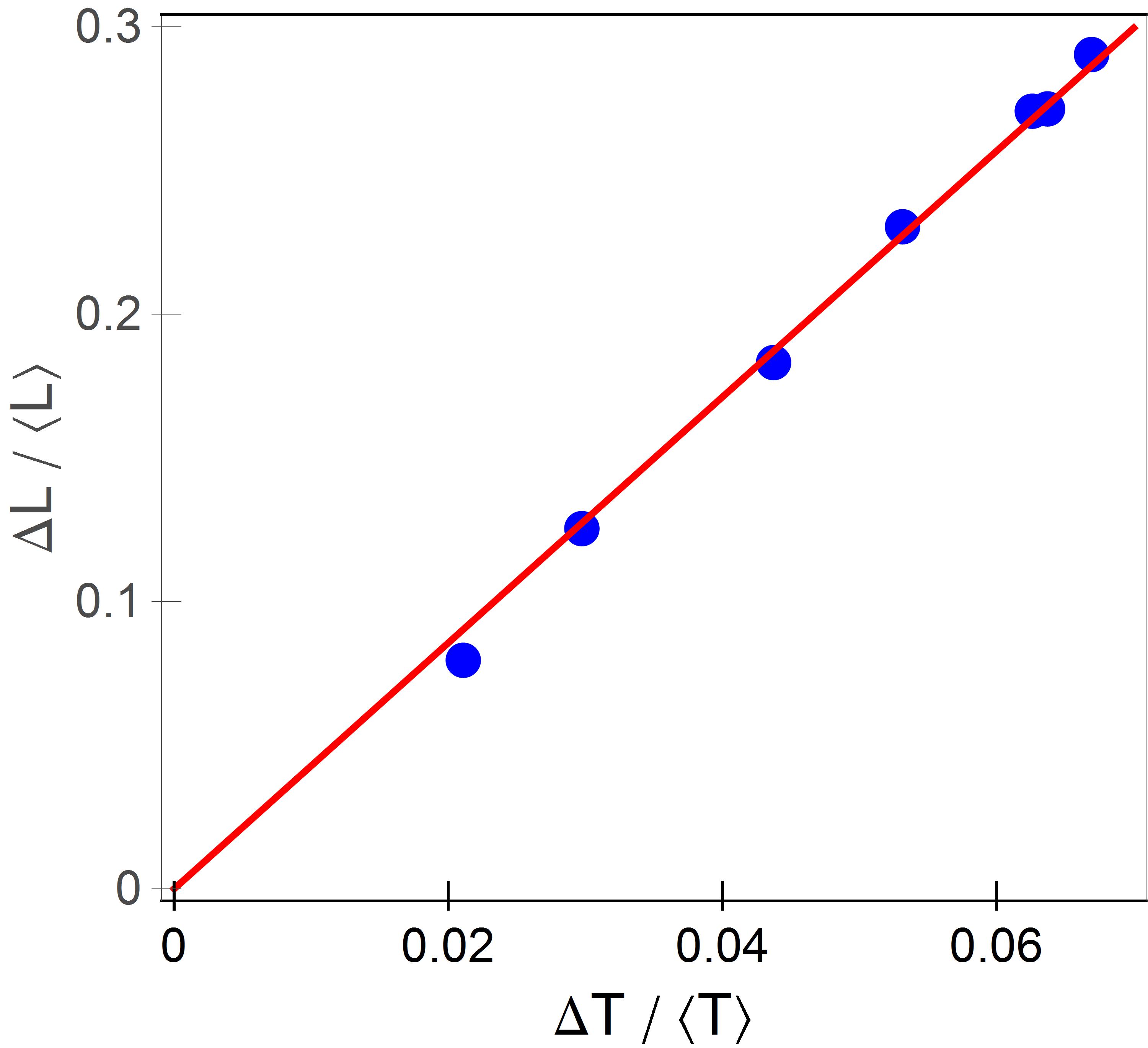,width=2.5in,height=2.2in,clip=5,angle=0}
\vspace*{-0.4cm}
\caption{ $\Delta T / \langle T\rangle$ {\it vs.}~$\Delta L / \langle L \rangle$
  in Pb+Pb collisions at $\sNN = 5.02$ TeV collision energy at various
  centralities~\cite{DREENAc,DREENAb}. The more peripheral the collision, the larger
  the values. The red solid line depicts linear fit to the values.}
\label{LTcorrelation}
\end{figure}

By combining  Eqs.~(\ref{RaaEstimate}) and~(\ref{v2approx}), we obtain:
\begin{eqnarray}
  \frac{v_{2}}{1-\RAA} \approx \left( \frac{b}{2} \frac{\Delta L}{\langle L \rangle}
                                   - \frac{a}{2}\frac{\Delta T}{\langle T \rangle} \right).
\label{v2RaaRatio}
\end{eqnarray}
This ratio carries information on the anisotropy of the system, but
through both spatial ($\Delta L / \langle L \rangle$) and temperature
($\Delta T / \langle T \rangle$) variables. From
Eq.~(\ref{v2RaaRatio}), we see the usefulness of the (approximate)
analytical derivations, since the term $(1-\RAA)$ in the denominator
could hardly have been deduced intuitively, or pinpointed by numerical
trial and error. Figure~\ref{LTcorrelation} shows a linear dependence
$\Delta L /\langle L\rangle \approx c \Delta T/\langle T\rangle$,
where $c \approx 4.3$, with the temperature evolution given by
one-dimensional Bjorken expansion, as sufficient to describe the early
evolution of the system. Eq.~(\ref{v2RaaRatio}) can thus be simplified
to
\begin{eqnarray}
  \frac{v_{2}}{1-\RAA} & \approx &
     \frac{1}{2} \left(b - \frac{a}{c}\right)
       \frac{\langle L_{out}\rangle - \langle L_{in} \rangle}
            {\langle L_{out}\rangle + \langle L_{in} \rangle}
   \approx 0.57 \varsigma ,                                    \nonumber \\
{\rm where} \, \; \;  \varsigma & = & \frac{\langle L_{out}\rangle - \langle L_{in} \rangle}
                                           {\langle L_{out}\rangle + \langle L_{in} \rangle}
\; \; \; {\rm and} \;\; \; \frac{1}{2} (b - \frac{a}{c}) \approx 0.57, \; \; \; \; \; \;
\label{AsymetryEq0}
\end{eqnarray}
when $a \approx 1.2$ and $b \approx 1.4$. Consequently, the asymptotic
behavior of observables $\RAA$ and $v_2$ is such that, at high $\pT$,
their ratio is dictated solely by the geometry of the fireball.
Therefore, the anisotropy parameter $\varsigma$ can be extracted from
the high-$p_\perp$ experimental data.

\begin{figure}
\epsfig{file=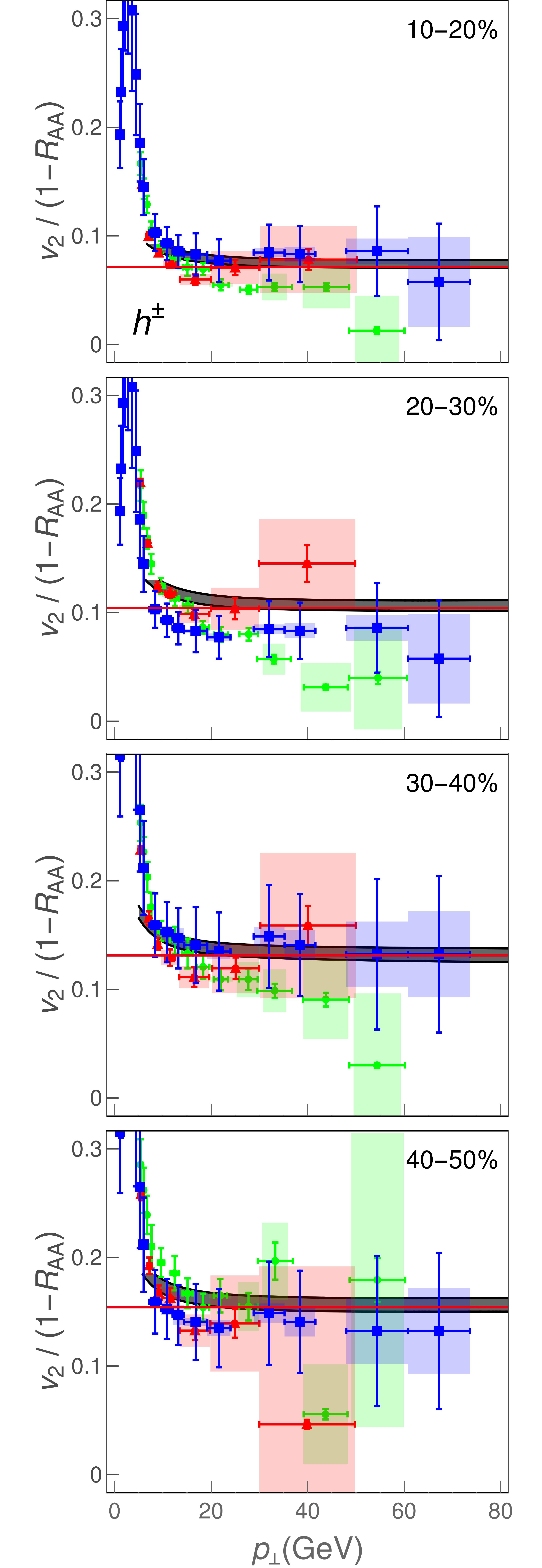,width=2.8in,height=7.8in,clip=5,angle=0}
\vspace*{-0.2cm}
\caption{Theoretical predictions for $v_{2}/(1-R_{AA})$ ratio of charged hadrons as a
  function of transverse momentum $p_{\perp}$ compared with $5.02$ TeV
  $Pb+Pb$ ALICE~\cite{ALICE_CH_RAA,ALICE_CH_v2} (red triangles),
  CMS~\cite{CMS_CH_RAA,CMS_CH_v2} (blue squares) and
  ATLAS~\cite{ATLAS_CH_RAA,ATLAS_CH_v2} (green circles) data. Panels
  correspond to 10-20\%, 20-30\%, 30-40\% and 40-50\% centrality
  bins. The gray band corresponds to the uncertainty in the magnetic
  to electric mass ratio $\mu_M/\mu_E$. The upper (lower) boundary of
  the band corresponds to $\mu_{M} / \mu_{E} = 0.4$
  ($0.6$)~\cite{Maezawa,Nakamura}. In each panel, the red line
  corresponds to the limit $0.57 \varsigma$ from
  Eq.~(\ref{AsymetryEq0}).}
\label{ExpComparisonFig}
\end{figure}

Regarding the parametrization used to derive Eq.~(\ref{AsymetryEq0})
(constants $a$, $b$ and $c$), we note that $a$ and $b$ are well
established within our dynamical energy loss formalism and follow from
$\RAA$ predictions that are extensively tested on experimental
data~\cite{NewObservable,MD_5TeV} and do not depend on the details of
the medium evolution. Regarding $c$, it may (to some extent) depend on
the type of the implemented medium evolution, but this will not affect
the obtained scaling, only (to some extent) the overall prefactor in
Eq.~(\ref{AsymetryEq0}).

{\it Numerical results:} To assess the applicability of the
analytically derived scaling in Eq.~(\ref{AsymetryEq0}), we calculate
$v_2/(1-\RAA)$ using our full-fledged numerical procedure for
calculating the fractional energy loss. This procedure is based on our
state-of-the-art dynamical energy loss formalism~\cite{MD_Dyn,MD_Coll},
which has several unique features in the description of high-$p_\perp$
parton medium interactions:
{\it i)} The formalism takes into account a {\it finite size, finite
  temperature} QCD medium consisting of {\it dynamical} (that is
moving) partons, contrary to the widely used static scattering
approximation and/or medium models with vacuum-like propagators
(e.g.~\cite{BDMPS,ASW,GLV,HT}).
{\it ii)} The calculations are based on the finite-temperature
generalized hard-thermal-loop approach~\cite{Kapusta}, in which the
infrared divergences are naturally
regulated~\cite{MD_Dyn,MD_Coll,DG_TM}.
{\it iii)} Both radiative~\cite{MD_Dyn} and collisional~\cite{MD_Coll}
energy losses are calculated under {\it the same} theoretical
framework, applicable to both light and heavy flavor.
{\it iv)} The formalism is generalized to the case of finite
magnetic~\cite{MD_MagnMass} mass, running coupling~\cite{MD_PLB} and
towards removing the widely used soft-gluon approximation~\cite{BDD_sga}.
The formalism was further embedded into our recently developed
DREENA-B framework~\cite{DREENAb}, which integrates initial momentum
distribution of leading partons~\cite{VitevDist}, energy loss with
path-length~\cite{ALICE_charm} and multi-gluon~\cite{GLV_suppress}
fluctuations and fragmentation functions~\cite{DSS}, in order to
generate the final medium modified distribution of high-$p_\perp$
hadrons. The framework was recently used to obtain joint $R_{AA}$
and $v_2$ predictions for $5.02$~TeV Pb+Pb collisions at the
LHC~\cite{DREENAb}, showing a good agreement with the experimental
data.

We have previously shown~\cite{Blagojevic_JPG} that all the model
ingredients noted above have an effect on the high-$\pT$ data, and
thus should be included to accurately explain it. In that respect,
our model is different from many other approaches, which use a
sophisticated medium evolution, but an (over)simplified energy loss
model. Our previous work, however, shows that, for explaining the
high-$p_\perp$ data, an accurate description of high-$p_\perp$
parton-medium interactions is at least as important as an advanced
medium evolution model. For example, the dynamical energy loss
formalism, embedded in 1D Bjorken expansion, explains well the $v_2$
puzzle~\cite{DREENAb}, i.e., the inability of other models to jointly
explain $R_{AA}$ and $v_2$ measurements. To what extent the dynamical energy
loss predictions will change when embedded in full
three-dimensional evolution is at the time of this writing still unknown,
but our previous results nevertheless make it plausible that
calculations employing simple one-dimensional expansion can provide
valuable insight into the behavior of jets in the medium.

Our results for longitudinally expanding system (1D Bjorken), and the
corresponding data are shown in Fig.~\ref{ExpComparisonFig}.  The gray
band shows our full DREENA-B result (see above) with the band
resulting from the uncertainty in the magnetic to electric mass ratio
$\mu_M/\mu_E$~\cite{Maezawa,Nakamura}. The red line corresponds to the
$0.57\varsigma$ limit from Eq.~(\ref{AsymetryEq0}), where $\varsigma$
is the anisotropy of the path lengths used in the DREENA-B
calculations~\cite{DREENAb,DREENAc}. Importantly, for each centrality,
the asymptotic regime -- where the $v_2/(1-\RAA)$ ratio does not
depend on $\pT$, but is determined by the geometry of the system -- is
already reached from $\pT \sim 20$--30 GeV; the asymptote corresponds
to the analytically derived Eq.~(\ref{AsymetryEq0}), within $\pm 5\%$
accuracy. It is also worth noticing that our prediction of asymptotic
behavior was based on approximations which are not necessarily valid
in these calculations, but the asymptotic regime is nevertheless
reached, telling that those assumptions were sufficient to capture the
dominant features. If, as we suspect, the high-$p_\perp$ parton-medium
interactions are more important than the medium evolution model in
explaining the high-$p_\perp$ data, this behavior reflects this
importance and the analytical derivations based on a static medium
may capture the dominant features seen in Fig.~\ref{ExpComparisonFig}.

Furthermore, to check if the experimental data support the derived
scaling relation, we compare our results to the
ALICE~\cite{ALICE_CH_RAA,ALICE_CH_v2}, CMS~\cite{CMS_CH_RAA,CMS_CH_v2}
and ATLAS~\cite{ATLAS_CH_RAA,ATLAS_CH_v2} data for $\sNN = 5.02$ TeV
$Pb+Pb$ collisions. The experimental data, for all three experiments,
show the same tendency, i.e., the independence on the $p_\perp$ and a
consistency with our predictions, though the error bars are still
large. Therefore, from Fig.~\ref{ExpComparisonFig}, we see that at
each centrality both the numerically predicted and experimentally
observed $v_{2}/(1-\RAA)$ approach the same high-$p_\perp$ limit.
This robust, straight line, asymptotic value carries information about
the system's anisotropy, which is, in principle, simple to infer from the
experimental data.

Ideally, the experimental data (here from ALICE, CMS and ATLAS) would
overlap with each other, and would moreover have small error bars. In
such a case, the data could be used to directly extract the anisotropy
parameter $\varsigma$ by fitting a straight line to the high-$\pT$
part of the $v_2/(1-\RAA)$ ratio. While such direct anisotropy
extraction would be highly desirable, the available experimental data
are unfortunately still not near the precision level needed to
implement this. However, we expect this to change in the upcoming
high-luminosity $3^{\mathrm{rd}}$ run at the LHC, where the error bars
are expected to be significantly reduced, so that this procedure can be
directly applied to experimental data.

It is worth remembering that the anisotropy parameter $\varsigma$,
which can be extracted from the high-$\pT$ data, is not the commonly
used anisotropy parameter $\epsilon_2$,
\begin{equation}
 \epsilon_2 = \frac{\langle y^2-x^2 \rangle}{\langle y^2+x^2 \rangle}
            = \frac{\int\dif x\,\dif y\, (y^2-x^2)\, \rho(x,y)}
                   {\int\dif x\,\dif y\, (y^2+x^2)\, \rho(x,y)},
 \label{eccentricity}
\end{equation}
where $\rho(x,y)$ is the initial density distribution of the QGP
droplet.  We may also expect, that once the transverse expansion is
included in the description of the evolution, the path-length
anisotropy $\varsigma$ reflects the time-averaged anisotropy of the
system, and therefore is not directly related to the initial-state
anisotropy $\epsilon_2$. Nevertheless, it is instructive to check how
the path-length anisotropy in our simple model relates to conventional
$\epsilon_2$ values in the literature. For this purpose we construct a
variable
\begin{equation}
  \epsilon_{2L} = \frac{\langle L_{out}\rangle^2 - \langle L_{in}\rangle^2}
                  {\langle L_{out}\rangle^2 + \langle L_{in}\rangle^2}
             = \frac{2 \varsigma}{1+\varsigma^2}.
  \label{asymmetryRelation}
\end{equation}
We have checked that for different density distributions $\epsilon_2$
and $\epsilon_{2L}$ agree within $\sim$10\% accuracy. 

\begin{figure}
\epsfig{file=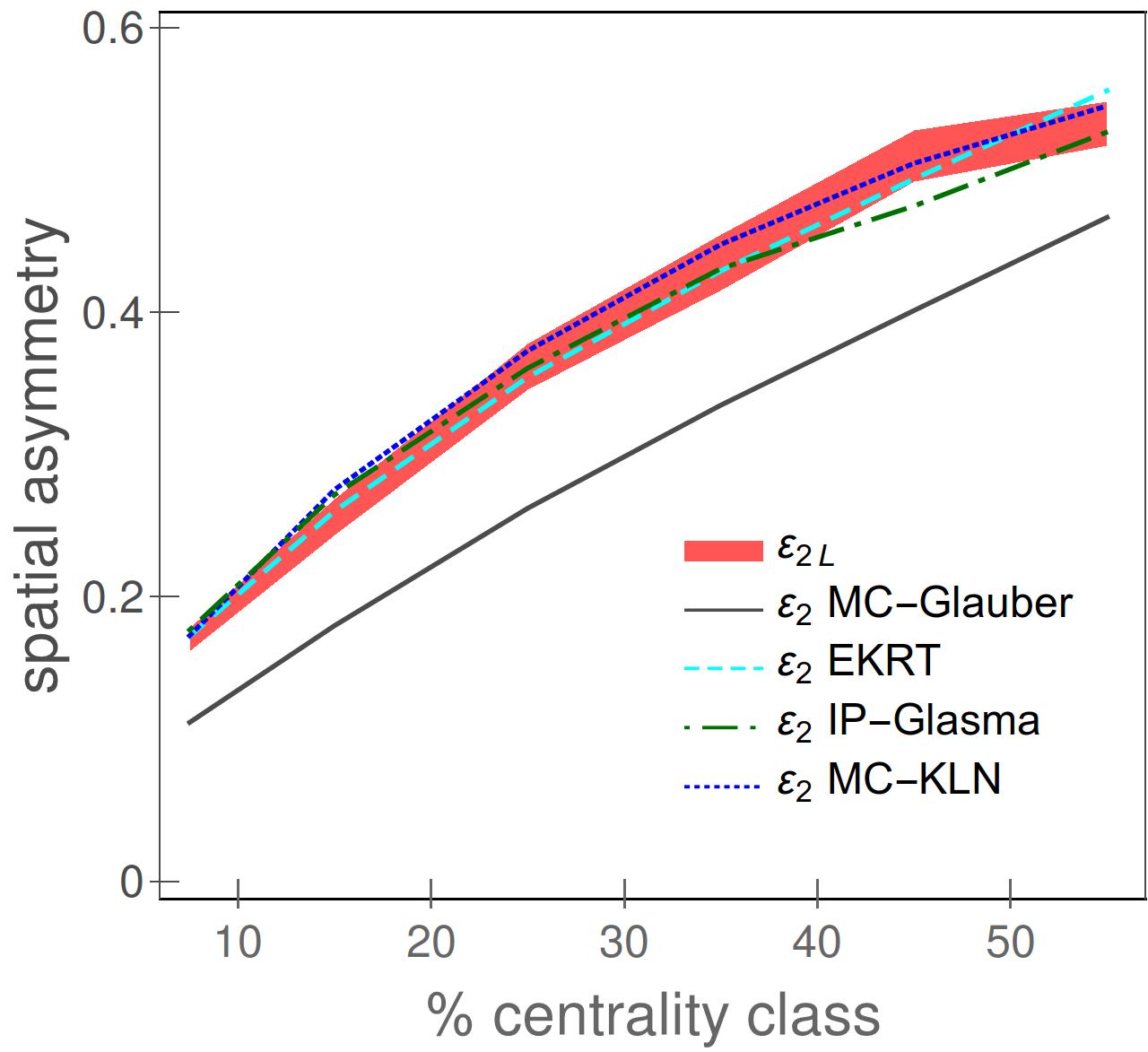,height=2.5in,clip=5,angle=0}
\vspace*{-0.2cm}
\caption{Comparison of $\epsilon_{2 L}$ (red band) obtained from our
  method, with $\epsilon_2$ calculated using Monte Carlo (MC)
  Glauber~\cite{MCGlauber} (gray band), EKRT~\cite{EKRT} (the purple
  band), IP-Glasma~\cite{IPGlasma} (green dot-dashed curve), and
  MC-KLN~\cite{MCKLN} (blue dotted curve) approaches. MC-Glauber and
  EKRT results correspond to 5.02 TeV, while IP-Glasma and MC-KLN
  correspond to 2.76 TeV $Pb+Pb$ collisions at the LHC.}
\label{epsilon2comparison}
\end{figure}

We have extracted the parameters $\varsigma$ from the DREENA-B results
shown in Fig.~\ref{ExpComparisonFig}; the corresponding $\epsilon_{2L}$
results are shown as a function of centrality in
Fig.~\ref{epsilon2comparison} and compared to $\epsilon_2$ evaluated
using various initial-state models in the
literature~\cite{MCGlauber,EKRT,IPGlasma,MCKLN}.  Note that
conventional (EKRT~\cite{EKRT}, IP-Glasma~\cite{IPGlasma})
$\epsilon_2$ values trivially agree with our {\it initial}
$\epsilon_2$ (not shown in the figure), i.e., the initial
$\epsilon_2$ characterize the anisotropy of the path lengths used as
an input to DREENA-B, which we had chosen to agree with the
conventional models~\footnote{Binary collision scaling calculated
  using optical Glauber model with additional cut-off in the tails of
  Woods-Saxon potentials, to be exact.}. It is, however, much less
trivial that, through this procedure, in which we calculate the ratio
of $v_2$ and $1-R_{AA}$ through full DREENA framework, our {\it
  extracted} $\epsilon_{2 L}$ almost exactly recovers our initial
$\epsilon_{2}$. Note that $\epsilon_{2}$ is {\it indirectly}
introduced in $R_{AA}$ and $v_2$ calculations through path-length
distributions, while our calculations are performed using full-fledged
numerical procedure, not just Eq.~(\ref{ElossEstimate}). Consequently,
such direct extraction of $\epsilon_{2 L}$ and its agreement with our
initial (and consequently also conventional) $\epsilon_{2}$ is highly
nontrivial and gives us a good deal of confidence that $v_2/(1-\RAA)$
is related to the anisotropy of the system only, and not its material
properties.

{\it Summary:} High-$p_\perp$ theory and data are traditionally used
to explore interactions of traversing high-$p_\perp$ probes with QGP,
while bulk properties of QGP are obtained through low-$p_\perp$ data
and the corresponding models. On the other hand, it is clear that
high-$p_\perp$ probes are also powerful tomography tools since they
are sensitive to global QGP properties. We here demonstrated this in
the case of spatial anisotropy of the QCD matter formed in
ultrarelativistic heavy-ion collisions. We used our dynamical energy
loss formalism to show that a (modified) ratio of two main
high-$p_\perp$ observables, $\RAA$ and $v_2$, approaches an asymptotic
limit at experimentally accessible transverse momenta, and that this
asymptotic value depends only on the shape of the system, not on its
material properties. However, how exactly this asymptotic value
reflects the shape and anisotropy of the system requires further study
employing full three-dimensional expansion, which is our current work in progress.
The experimental accuracy does not yet allow the extraction of the
anisotropy from the data using our scheme, but once the accuracy
improves in the upcoming LHC runs, we expect that the anisotropy of the
QGP formed in heavy-ion collisions can be inferred directly from the
data. Such an experimentally obtained anisotropy parameter would provide
an important constraint to models describing the early stages of
heavy-ion collision and QGP evolution, and demonstrate synergy of
high-$p_\perp$ theory and data with more common approaches for
inferring QGP properties.

{\em Acknowledgments:}
We thank Jussi Auvinen, Hendrik van Hees, Etele Molnar and Dusan Zigic
for useful discussions. We also thank Tetsufumi Hirano and Harri Niemi
for sharing their MC-KLN and EKRT results with us. This work is
supported by the European Research Council, Grant No.~ERC-2016-COG:~725741,
and by the Ministry of Science and Technological Development of the
Republic of Serbia, under Projects No.~ON171004 and No.~ON173052.

\end{document}